\documentclass[american,aps,prl,reprint,superscriptaddress,longbibliography]{revtex4-1}

\usepackage{amsmath,amssymb}

\usepackage{bbm}
\usepackage{color}
\usepackage{overpic}
\usepackage[american]{babel}
\usepackage[utf8]{inputenc}
\usepackage{times}
\usepackage{epstopdf}
\usepackage{braket}    %Dirac-Notation
\usepackage{amsthm}

\newcommand{\ketbra}[2]{|#1\rangle\!\langle#2|}

\definecolor{mygrey}{gray}{0.35}
\definecolor{myblue}{rgb}{0.2,0.2,0.8}
\definecolor{myzard}{cmyk}{0,0,0.05,0}
\definecolor{mywhite}{rgb}{1,1,1}
\definecolor{myred}{rgb}{0.9,0.1,0}
\usepackage[colorlinks=true,citecolor=myblue,linkcolor=myblue,urlcolor=myblue]{hyperref}

\newtheoremstyle{customStyle1}    %name of the style tobe used
{0pt}                             %measure of space to leave above the theorem. E.g.: 3pt
{0pt}                             %measure of space to leave below the theorem. E.g.: 3pt
{\normalfont}                     %name of font to use in the body of the theorem
{\parindent}                      %measure of space to indent
{\em}                             %name of head font
{. --}                            %punctuation between head and body
{.5em}                            %space after theorem head
{\thmname{#1}\thmnumber{ #2}\thmnote{ (#3)}}   % Manually specify head

\ifx\proof\undefined
\newenvironment{proof}[1][\protect\proofname]{\par
        \normalfont\topsep6\p@\@plus6\p@\relax
        \trivlist
        \itemindent\parindent
        \item[\hskip\labelsep\scshape #1]\ignorespaces
}{%
\endtrivlist\@endpefalse
}
\providecommand{\proofname}{Proof}
\fi

\newtheorem*{theorem*}{Theorem}
\newtheorem*{lemma*}{Lemma}
\newtheorem*{corollary*}{Corollary}
\newtheorem*{observation*}{Observation}
\newtheorem*{proposition*}{Proposition}

\newcommand{\tr}{\operatorname{\bf{tr}}}                  % trace of a matrix
\newcommand{\id}{\mathbbm{1}}

\let\oldforall\forall
\renewcommand{\forall}{\quad \oldforall}

\begin{document}
\title{Experimental quantification of coherence of a tunable quantum detector}%in quantum measurements}

\author{Huichao Xu}
\thanks{These three authors contributed equally}
\affiliation{National Laboratory of Solid State Microstructures,
Key Laboratory of Intelligent Optical Sensing and Manipulation, College of Engineering and Applied Sciences
and Collaborative Innovation Center of Advanced Microstructures, Nanjing University, Nanjing 210093, China}

\author{Feixiang Xu}
\thanks{These three authors contributed equally}
\affiliation{National Laboratory of Solid State Microstructures,
Key Laboratory of Intelligent Optical Sensing and Manipulation, College of Engineering and Applied Sciences
and Collaborative Innovation Center of Advanced Microstructures, Nanjing University, Nanjing 210093, China}

\author{Thomas Theurer}
\thanks{These three authors contributed equally}
\affiliation{Institute of Theoretical Physics and IQST, Universität Ulm, Albert-Einstein-Allee
        11, D-89069 Ulm, Germany}

\author{Dario Egloff}
\affiliation{Institute of Theoretical Physics and IQST, Universität Ulm, Albert-Einstein-Allee
	        11, D-89069 Ulm, Germany}
\affiliation{Technical University Dresden, Institute of Theoretical Physics, D-01062 Dresden, Germany}

\author{Zi-Wen Liu}
\affiliation{Perimeter Institute for Theoretical Physics, Waterloo, ON N2L 2Y5, Canada}
\affiliation{Center for Theoretical Physics, Massachusetts Institute of Technology, Cambridge,
             Massachusetts 02139, USA}

\author{Nengkun Yu}
\affiliation{Centre for Quantum Computation and Intelligent Systems, Faculty of Engineering and
              Information Technology, University of Technology Sydney, New South Wales 2007,
	      Australia}

\author{Martin B. Plenio}
\email[]{martin.plenio@uni-ulm.de}
\affiliation{Institute of Theoretical Physics and IQST, Universität Ulm, Albert-Einstein-Allee
        11, D-89069 Ulm, Germany}

\author{Lijian Zhang}
\email[]{lijian.zhang@nju.edu.cn}
\affiliation{National Laboratory of Solid State Microstructures,
Key Laboratory of Intelligent Optical Sensing and Manipulation, College of Engineering and Applied Sciences
and Collaborative Innovation Center of Advanced Microstructures, Nanjing University, Nanjing 210093, China}

\begin{abstract}
  Quantum coherence is a fundamental resource that quantum technologies exploit to achieve performance
  beyond that of classical devices. A necessary prerequisite to achieve this advantage is the ability of measurement devices to detect  coherence from the measurement statistics. Based on a recently developed resource theory of quantum operations, here we quantify experimentally the ability of a typical quantum-optical detector, the weak-field
  homodyne detector, to detect coherence. We derive an improved algorithm for quantum detector tomography and
  apply it to reconstruct the positive-operator-valued measures (POVMs) of the detector in different configurations. The reconstructed POVMs are then employed to evaluate how well the detector can detect coherence using two computable measures. As the first experimental investigation of quantum measurements from a resource theoretical perspective, our work sheds new light on the rigorous evaluation of the performance of a quantum measurement apparatus.
\end{abstract}
\maketitle

{\em Introduction.} -- Quantum coherence plays an indispensable role in quantum technologies including, for example, quantum computation \cite{PhysRevA.93.012111,PhysRevA.95.032307}, quantum coding \cite{Wiesner83} and key distribution \cite{BB.84}, quantum metrology \cite{PhysRevLett.116.120801,1751-8121-51-2-025302} and quantum biology \cite{Lloyd_2011,huelga2013vibrations}. Therefore, the quantitative assessment of quantum coherence as a resource has
attracted widespread interest \cite{arXiv:quant-ph/0612146,PhysRevLett.113.140401,PhysRevLett.116.120404,Theurer2017,RevModPhys.89.041003}. Until recently, most of the research concerned with the assessment of quantum coherence as a resource
focused on the coherence in quantum states
(see Ref.~\cite{RevModPhys.89.041003}
for a review). Following approaches in the resource
theory of entanglement \cite{PhysRevA.62.052317,PhysRevA.62.030301}, the coherence properties of quantum operations
have also begun to be examined by their ability to create or increase coherence in quantum states
\cite{PhysRevA.92.032331,GarciaDiazEP16,PhysRevA.95.052306,PhysRevA.95.052307}.

However, to exploit quantum coherence for different applications, it is generally insufficient
to only create and manipulate coherence: we also must be able to detect coherence in the sense that its presence makes a difference
in measurement statistics~\cite{PhysRevX.6.041028,PhysRevLett.118.060502,Smirne_2018,Theurer2018Quantifying,milz2019non}.
To quantify how well a measurement can detect coherence, a theoretical framework in the form of a resource theory
on the level of operations has been proposed~\cite{Theurer2018Quantifying}, allowing to address this question rigorously (see Refs.~\cite{liu2019operational,arXiv:1904.04201,gour2019quantify,gour2019entanglement,buml2019resource,wang2019resource,xu2019coherence} for related work). Other approaches connecting measurements with quantum coherence and resource theories were recently presented in Refs.~\cite{bischof2018resource,guff2019measurement,PhysRevResearch.1.033020}.

Here and in the following, the term measurement refers to the whole measurement apparatus, which is treated as a black box that takes quantum states as input and outputs a classical signal.
Hence a measurement is fully characterized by its positive-operator-valued measure (POVM).
Yet, if we want to know how well a measurement can detect coherence, we cannot simply refer to the POVM, as it contains too much information. Instead, we want to obtain a single number that describes the measurement's performance and, in particular, allows to compare it to the other measurements. Such a comparison can only be sensible if we establish a well-motivated notion of what it means that one measurement is better suited for detecting coherence than another one, which is reflected in the numbers associated to the measurements.
Resource theories are precisely developed for this task.
From physically motivated constraints, these mathematical frameworks derive resource measures and their evaluation extracts, e.g., the usefulness of the POVM in terms of the desired single number. This systematic and objective approach distinguishes measures derived within a resource theory from ad hoc measures and is one of the reasons for the popularity of resource theories.

Using the resource theory presented in Ref.~\cite{Theurer2018Quantifying},
here we quantify experimentally the capability of a
measurement apparatus, the weak-field homodyne detector (WFHD)~\cite{PhysRevLett.102.080404,Puentes_2010}, to detect coherence. In contrast
to photon-counting detectors that are sensitive to the intensity or particle behavior of input light fields only~\cite{hadfield2009single}, the WFHD
mixes the input light field with a phase-reference field, the local oscillator (LO), at a beam splitter and thus is able to measure the wave-like properties of the input field~\cite{donati2014observing}. In this sense, both the LO and the beam splitter are part of the WFHD. The difference between a photon counting detector and the WFHD can be further revealed by the matrix representations of their POVMs in photon-number basis: the former is completely diagonal and hence incoherent, while the latter has off-diagonal matrix elements and provides sensitivity to the coherence of the input states. Moreover, the WFHD can be tuned to interpolate continuously between photon-counting (incoherent) measurements and phase-dependent measurements by adjusting the intensity of the LO~\cite{PhysRevA.101.031801}. This detector has found important applications in not only state detection
\cite{PhysRevLett.105.253603,olivares2018homodyne,allevi2017homodyne} and state discrimination~\cite{nphoton.2012.316,nphoton.2014.280}, but
also for state preparation~\cite{bimbard2010quantum,  PhysRevA.81.042109, Akhlaghi:11,Yukawa:13,andersen2015hybrid}.

{\em Detecting coherence.}--
To quantify how well a measurement can detect coherence, we use the methods developed in Ref.~\cite{Theurer2018Quantifying}.
In the following, we shortly review the parts relevant for this work but refer to the original
paper for details. A quantum state $\rho$ is called incoherent with respect to a fixed orthonormal basis
$I=\{\ket{i}\}$ if $\rho$ is a statistical mixture of elements of $I$, i.e., $\rho=\sum_i c_i \ketbra{i}{i}$.
All other states are coherent. The total dephasing operation $\Delta$, which is defined by
\begin{equation}
	\Delta(\rho) = \sum_i \ketbra{i}{i}\rho\ketbra{i}{i},
	\label{eq:dephasing}
\end{equation}
is a resource destroying map~\cite{PhysRevLett.118.060502}, i.e., its output is always incoherent and incoherent
states are invariant under its action. Now consider a POVM $\{\Pi_n:\Pi_n\ge0, \sum_n \Pi_n=\id\}$, where $\id=\sum_j\ketbra{j}{j}$ is the identity operator and $\Pi_n=\sum_{j,k}\pi_n^{j,k}\ketbra{j}{k}$ is the POVM element associated to outcome $n$. Such a POVM cannot detect
coherence if its measurement statistics $p_n=\tr(\rho\Pi_n)$ is independent of the coherence within the input state, i.e.,
\begin{equation}
	\tr(\rho \Pi_n) = \tr(\Delta({\rho}) \Pi_n) \ \forall \rho,n.
	\label{eq:incoherent_measure}
\end{equation}
This result implies that for an incoherent measurement every $\Pi_n$ is diagonal in $I$~\cite{Theurer2018Quantifying}, while if $\Pi_n$ has off-diagonal elements the measurement will be able to detect coherence.

Storing measurement outcomes
in the incoherent basis of an auxiliary system, every quantum measurement can be represented by a quantum channel,
e.g., a POVM $\{\Pi_n\}$ by
\begin{align}
	\Theta(\rho)=\sum_n \tr(\rho\Pi_n) \ketbra{n}{n}.
\end{align}
Treating subselection this way, a general quantum operation $\Phi$ cannot detect coherence iff
\begin{align}
	\Delta \Phi =\Delta \Phi \Delta.
\end{align}
The set of these detection-incoherent operations is denoted by $\mathcal{DI}$. This allows us to present two
well defined and computable functionals quantifying how well
an operation can detect coherence~\cite{Theurer2018Quantifying}; the diamond measure
\begin{equation}
	M_\diamond(\Theta)=\min_{\Phi \in \mathcal{DI}} \|\Delta (\Theta - \Phi) \|_\diamond
	\label{eq:diamond}
\end{equation}
and the NSID measure,
\begin{equation}
	\tilde{M}_\diamond(\Theta)=\min_{\Phi\in \mathcal{DI}} \|\Delta \left(\Theta- \Phi \right) \|_1.
	\label{eq:NSID}
\end{equation}
It is worthwhile to mention that the NSID measure is directly related to the success probability of simulating $\Theta$ by operations that cannot
detect coherence. Furthermore, the diamond measure provides an upper bound on the average coherence that
can be prepared remotely when the measurement is applied on one part of the maximally entangled bipartite state  (see Sec.~V of the Supplemental Material (SM)~\cite{Supplementary_material} for details). The coherence of a quantum measurement can be evaluated in
two steps: map the measurement to a trace-preserving operation, and then calculate the coherence of the
operation using Eq.~(\ref{eq:diamond}) or Eq.~(\ref{eq:NSID}). While these measures are generally different,
remarkably, for channels with output dimension two (two measurement outcomes) we have been able to prove
that the two measures coincide (see Sec.~VI of the SM~\cite{Supplementary_material} for details).

\begin{figure}[t]
	\centering
	\includegraphics[width=0.45\textwidth]{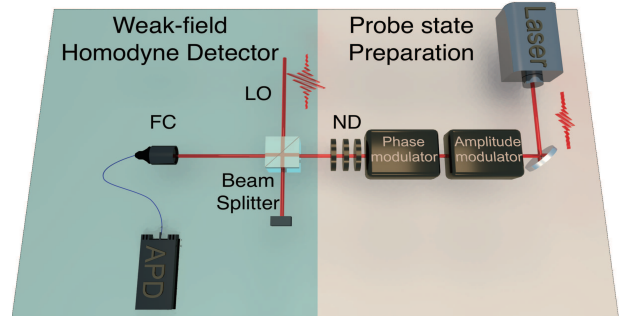}
	\caption[width=1\textwidth]{Schematic diagram of the experimental setup. The experimental setup can be divided into two parts, the weak-field homodyne detector (WFHD) part (left) and the probe state preparation part (right). The output of a laser goes through an amplitude modulator and a phase modulator followed by calibrated neutral density filters (ND) to prepare a set of coherent states as the probe states. The probe state interferes with the local oscillator (LO) at a 50:50 beam splitter (BS) with one output mode coupled into a single mode fiber and detected by an avalanche photodiode (APD). FC denotes a fiber coupler.}
	\label{fig:exp}
\end{figure}

\begin{figure*}[ht]
	\centering
	\includegraphics[width=1\textwidth]{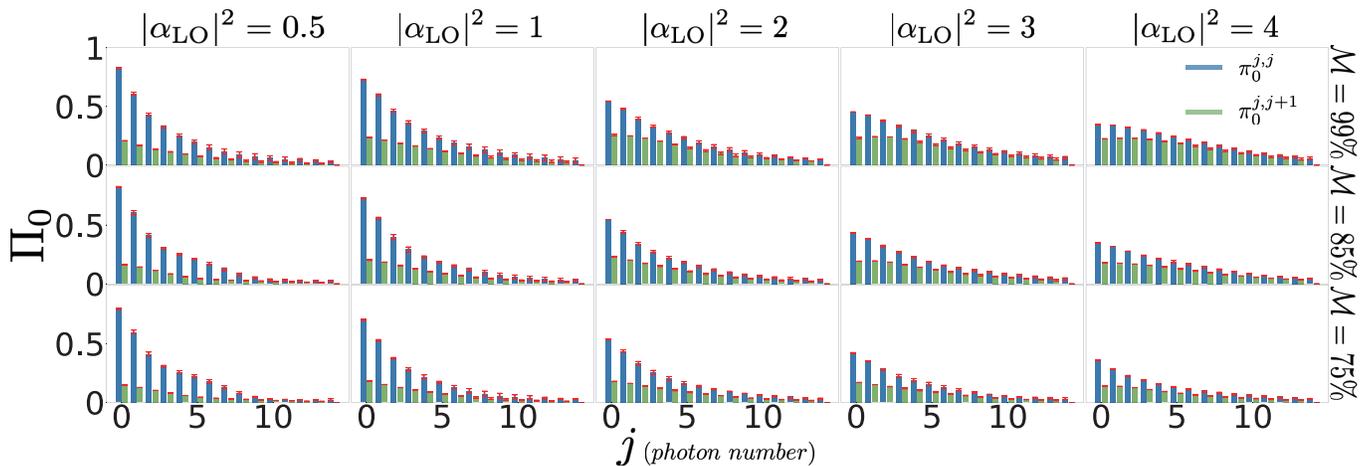}
	\caption[width=1\textwidth]{Experimentally reconstructed no-click POVM elements $\Pi_0$ of the weak-field homodyne detector with different LO intensities $|\alpha_{\mathrm{LO}}|^2$ and mode overlaps $\mathcal{M}$. The reflectivity of the beam splitter is 0.5 and the quantum efficiency of the APD is 59\%. The POVMs are reconstructed up to 70 photons and shown up to 15 photons in the figure. Only the diagonal ($\pi_0^{j,j}=\bra{j}\Pi_0\ket{j}$, blue bars) and the first off-diagonal ($\pi_0^{j,j+1}=\bra{j}\Pi_0\ket{j+1}$, green bars) of the POVM elements are shown here. The error bars originate from the fluctuations in the preparation of the probe states used for tomography.
			}
	\label{fig:off}
\end{figure*}
{\em Experimental setup.}-- The quantum detector we investigate here is a weak-field homodyne detector (WFHD) which is tunable with various parameters. Similar to a standard homodyne detector, the WFHD combines the input state with a coherent optical field $|\alpha_{\mathrm{LO}}\rangle$, the LO. Yet the intensity of the LO $|\alpha_{\mathrm{LO}}|^2$ in WFHD is reduced to low photon numbers. Therefore, instead of photodiodes, a photon-counting detector, an avalanche photodiode (APD) is used to detect the interference signal. Since the LO acts as a phase reference, a WFHD is a phase sensitive detector, whose properties have been well studied in \cite{PhysRevLett.102.080404,Puentes_2010,donati2014observing}.

In this work we study the coherence in a WFHD, as shown in Fig. \ref{fig:exp}, under various configurations. Since an APD is a binary detector, there are two outcomes of the detector: no-click (0) and click (1). We fix the ratio of the beam splitter as 50:50, and set the average photon number of the LO $|\alpha_{\mathrm{LO}}|^2$ to five different values 0.5, 1, 2, 3 and 4. For each LO intensity, the degree of the mode overlap between the LO and the input state is chosen to be $\mathcal{M} = 99\%, 85\%, 75\%$. Due to the relatively complex theoretical model of the detector and the experimental imperfections that are difficult to calibrate accurately, we apply quantum detector tomography~\cite{Lundeen2008Tomography,Feito2009Measuring,PhysRevLett.124.040402} to characterize experimentally the detector for different configurations. Quantum detector tomography (QDT) allows to reconstruct the POVM of an arbitrary quantum detector from the outcome statistics in response to a set of tomographyically complete probe states. In this work we use a set of coherent states $|\alpha\rangle$ as the probe states, which can be generated by modulating the amplitude and phase of the output of a laser. The probe states interfere with the LO at a 50:50 beam splitter with one output mode coupled into a single mode fiber for APD detection. The other output mode can be used for tracking the relative phase between probe and LO states. More details of the experimental setup can be found in the SM~\cite{Supplementary_material}.

{\em Experimental results.}--
We adopt a two-step method to quantify the coherence of the WFHD: first we reconstruct the POVM, which will then allow us to evaluate  Eq.~(\ref{eq:diamond}) or Eq.~(\ref{eq:NSID}) using numerical methods~\cite{Theurer2018Quantifying}. In principle, recording the statistics of the measurement outcomes for different probe states, the POVM can be estimated by inverting a set of linear equations given by the Born rule. However, taking into account experimental imperfections and statistical fluctuations, the POVM is usually reconstructed using a constrained convex optimization program. Here, we follow this approach and reconstruct the POVMs using an improved QDT method based on~\cite{Zhang_2012,Zhang2012Mapping}. We truncate the Hilbert space at the photon number of 70 which is sufficient to saturate the detector and reconstruct up to the fifth leading diagonals of the POVM elements. The details are given in the SM~\cite{Supplementary_material}.

Before moving forward to quantitatively evaluate the coherence of the POVMs with the diamond measure and the NSID measure, we first compare the POVMs associated to the different configurations of the WFHD. The reconstructed POVM elements of the no-click outcome are given in Fig.~\ref{fig:off}. We only present the diagonals (blue bars) and first off-diagonals (green bars) to elucidate the difference between different POVMs since they are the most significant ones. The three rows (from top to bottom) correspond to three different degrees of mode overlap between the input states and the LO, $\mathcal{M} = 99\%, 85\%, 75\%$, respectively. The five columns (from left to right) represent the five different average photon numbers of the LO used ($|\alpha_{\mathrm{LO}}|^2$ from 0.5 to 4). The complete POVMs are given in the SM~\cite{Supplementary_material}.

The off-diagonal elements of the density matrix determine the coherence of a quantum state \cite{PhysRevLett.113.140401}.
Recalling that the matrices of an incoherent POVM are completely diagonal in the incoherent basis, the off-diagonal part of the POVM elements also plays an important role in quantifying the coherence of a measurement. Indeed the diamond measure is bounded from below by the off-diagonal part of the POVM elements~\cite{Supplementary_material}. Therefore, we start the discussion by focusing on the off-diagonal parts of the POVM matrices. For a fixed LO intensity, it can be seen in Fig.~\ref{fig:off} that the off-diagonal elements decrease with the reduction of the mode overlap $\mathcal{M}$ between the input state and the LO, which suggests a reduced capability to detect the coherence of input states. This result can be understood by dividing the LO into two parts, one that overlaps with the input state and the other with the mode orthogonal to that of the input state~\cite{Laiho_2009}. The intensities of the two parts are
 $\mathcal{M}|\alpha_{\mathrm{LO}}|^2$ and $(1-\mathcal{M})|\alpha_{\mathrm{LO}}|^2$ respectively. Only the first part can interfere with the input state and provide a phase reference, resulting in the off-diagonal elements and the capability to detect coherence. The second part plays the role of background noise that can lead to decoherence of the quantum detectors~\cite{PhysRevLett.107.050504}. Therefore, a decrease in the mode overlap implies not only a reduced phase reference but also an increased background noise. The overall effect is a reduced sensitivity to coherence in the input state. In the limit of no mode overlap at all, the WFHD becomes an intensity detector with the LO acting as background noise, leading to a complete loss of its ability to detect coherence.

For a fixed mode overlap, we should distinguish between two cases. For near perfect mode overlap $\mathcal{M}=99\%$, all the LO interferes with the input state. A higher LO intensity implies a better phase reference and therefore larger off-diagonals. With increasing LO intensity, the peak of the off-diagonal elements also shifts to higher photon numbers, which can be understood by considering that interference between two beams shows maximal visibility when they have the same intensity.
In the case of the non-unit mode overlap, again the LO plays a dual role: as phase reference (with the intensity $\mathcal{M}|\alpha_{\mathrm{LO}}|^2$) and as phase-independent background noise (with the intensity $(1-\mathcal{M})|\alpha_{\mathrm{LO}}|^2$). Both effects increase with the intensity of the LO. The competition between the coherent interference and the incoherent background noise explains why the off-diagonals increase first and then decrease with the increase in the LO intensity for a non-unit mode overlap.

Now, we are ready to move on to the second step, evaluating the exact value of the coherence contained in WFHD. The diamond measure as given in Eq.~(\ref{eq:diamond}) and the NSID measure in Eq.~(\ref{eq:NSID}) are calculated. Based on the analysis in \cite{Theurer2018Quantifying} and the references therein, the diamond measure can be calculated efficiently using a semidefinite program. The evaluation of the NSID measure is more cumbersome but, as mentioned above, has a clearer operational meaning. In our case, however, the two measures coincide (see the SM~\cite{Supplementary_material} for the proof, where we also show that both measures are robust against errors in the reconstructed POVM elements). This allows us to use the efficient semidefinite program to evaluate both measures.

\begin{figure}[t]
	\centering
  \includegraphics[width=0.48\textwidth]{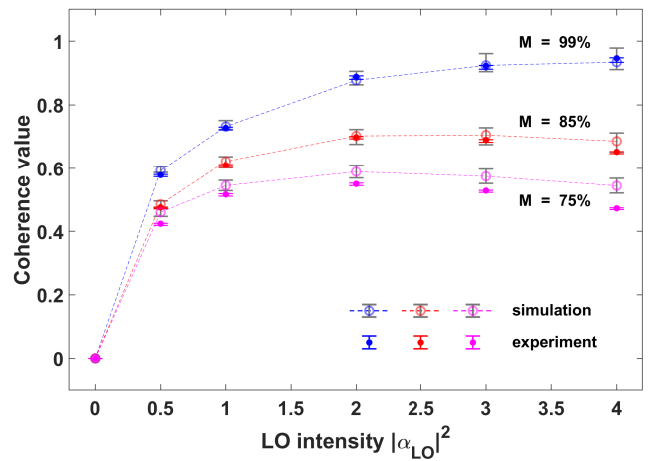} \caption[width=1\textwidth]{Diamond measure and non-stochasticity in detection (NSID) measure for the POVM elements of a weak-field homodyne detector with three mode overlaps $\mathcal{M}=99\%~ \text{(blue)},\ 85\%~ \text{(red)},\ 75\%~ \text{(magenta)}$.
  For each value of $\mathcal{M}$, we present the measures for different LO intensities $|\alpha_{\mathrm{LO}}|^2 = 0.5, 1, 2, 3, 4$. The circle symbols with gray error bars are linked by segment lines representing simulation results and the dot symbols with error bars are obtained from experimentally reconstructed POVMs. The diamond measure and the NSID measure coincide.}
	\label{fig:co}
\end{figure}

The coherence of the experimentally reconstructed POVMs of the WFHD are shown in Fig.~\ref{fig:co} by dots and the error bars originate again from the fluctuations in the intensities and phases of the probe states.
The measures are presented for three values of the mode overlap $\mathcal{M}=99\%$ (blue), $85\%$ (red),  $75\%$ (magenta) and five different intensities of the LO $|\alpha_{\mathrm{LO}}|^2 = 0.5, 1, 2, 3, 4$ per mode overlap. When the intensity of the LO is zero, the WFHD is degenerated into an APD with additional 50\% loss at the BS and the coherence is zero, which is also shown in the figure.

For comparison, we also simulated the POVM elements by numerically generating the statistics of the measurement outcomes with the configuration parameters of the WFHD in the experiment, then reconstructing the POVM using the simulated data.  The corresponding results are shown by circles with gray error bars which are linked by segmented lines for fixed mode overlap to show the tendency on the LO intensity clearly. In these simulations, we used experimentally determined parameters which themselves are inaccurate. The mode overlap is determined with an inaccuracy of up to 2\%, which is due to the limited precision of our power meter leading to the gray error bars. All in all, the experimental results match the simulations well.
The remaining discrepancy can be traced back to noise and fluctuations not taken into consideration in our simulations~\cite{Supplementary_material}. As expected, this decreases the coherence of the detector further and is an argument in favor of QDT, where such effects are taken into account automatically.

The estimated coherence shows a change similar to that in the off-diagonal elements of the POVM. For a fixed intensity of the LO, it is obvious that the coherence decreases with the reduction in the mode overlap, which agrees with the above analysis based on the POVM elements. For a fixed mode overlap, the relation between the coherence and the LO intensity is less obvious and can be explained with the same arguments we used when discussing the off-diagonals of the POVM elements. When the mode overlap is nearly perfect ($\mathcal{M}=99\%$), increasing LO intensity grants a better phase reference, leading to higher coherence.  In case of imperfect mode overlap, the dual role of the LO as both background noise and phase reference makes the connection between coherence and LO intensity more subtle: as we can see, the coherence first increases with increasing LO intensity and then decreases. The value of the coherence of a two-outcome measurement is bounded from above by 1~\cite{Theurer2018Quantifying}, which we nearly reach with increased LO amplitudes and perfect mode overlap.

{\em Conclusions and outlook.}--
Detecting coherence, a quantum resource at the core of nonclassical effects such as entanglement, is a necessary prerequisite to its exploitation in quantum technologies~\cite{PhysRevX.6.041028,PhysRevLett.118.060502,Smirne_2018}. It is therefore crucial to have detectors that can measure coherence and moreover, to know their performance precisely.
In this work, we have shown how one can use resource theories to study the performance of a relevant quantum detector, namely the tunable weak-field homodyne detector (WFHD). We develop an improved method of quantum detector tomography to reconstruct the positive-operator-valued measures (POVMs) describing the WFHD with different configurations. Using the reconstructed POVMs and two well-defined resource measures, we show how different parameters quantitatively affect the WFHD's capability to detect coherence.

Hence, this work presents the first rigorous experimental and theoretical analysis of one of the main nonclassical aspects, coherence, of quantum operations and detectors in particular. The good agreement between simulation and experiment demonstrates that abstract resource theories are applicable to practical, high-dimensional devices with sufficient precision. Therefore, our work bridges the gap between theory and experiment and provides quantitative tools for assessing improved designs of devices exploiting quantum effects.

\begin{acknowledgments}
	We thank Ish Dhand for helpful discussions. This work was supported by the National Key Research and Development Program of China under (Grant Nos. 2018YFA0306202 and 2017YFA0303703), the National Natural Science Foundation of China (Grant Nos. 11690032, 61490711, 11474159 and 11574145) and Fundamental Research Funds for the Central
Universities (Grant No. 020214380068). T.T., D.E., and M.B.P. acknowledge financial support by the ERC Synergy Grant BioQ (grant no 319130)
	and the state of Baden-Württemberg through bwHPC
		and the German Research Foundation (DFG) through grant no INST 40/467-1 FUGG (JUSTUS cluster). Numerical calculation were done with the help of Refs.~\cite{Sturm1999,Lofberg2004}.
\end{acknowledgments}

%\bibliography{reference_v15}


\end{document}